# Cell-like space charge configurations formed by self-organization in laboratory

**Erzilia Lozneanu** and **Mircea Sanduloviciu**
Department of Plasma Physics
Complexity Science Group
Al. I. Cuza University
6600 Iasi, Romania
msandu@uaic.ro

A phenomenological model of self-organization explaining the emergence of a complexity with features that apparently satisfy the specific criteria usually required for recognizing the appearance of life in laboratory is presented. The described phenomenology, justified by laboratory experiments, is essentially based on local self-enhancement and long-range inhibition. The complexity represents a primitive organism self-assembled in a gaseous medium revealing, immediately after its "birth", many of the prerequisite features that attribute them the quality to evolve, under suitable conditions, into a living cell.

## 1. Introduction

In this paper we would like to report on the possibility to create in laboratory a **c**omplex **s**pace **c**harge **c**onfiguration (CSCC) representing, in our opinion, the simplest possible system able to reveal operations usually attributed to a biological cell. It appears in a cold physical plasma, *i.e.* a medium presumable similar to those existent under prebiotic Earth's conditions, when an electrical spark creates a well-located nonequilibrium plasma. In spite of its gaseous nature, such a CSCC satisfies to a large extend the criteria usually required to recognize the creation of life in laboratory. Thus, similar to biological cells, the boundary of a self-assembled CSCC provides a selective confinement of an environment that qualitatively differs from the surrounding medium. The boundary appears as a spherical self-consistent electrical **d**ouble **l**ayer (DL) able to sustain and control operations such as: (i) capture and transformation of energy, (ii) preferential and rhythmic exchange of matter across the system boundary and internal transformation of matter by the means of a continuous "synthesis" of all components of the system. After its formation, the CSCC is able to replicate, by division, and to emit and receive information.

As proved by experimental and simulation studies [1-4] self-organization occurs in an intermittent fashion (intermittent self-organization) or in a stepwise fashion (cascading self-organization). Intermittent self-organization occurs when the system is gradually driven away from equilibrium by continuous injection of matter and energy, while cascading self-organization occurs when matter and energy are suddenly injected into the system so that it relaxes stepwise towards a minimum energy state.
Because the succession of physical processes involved in the intrinsically nonlinear mechanism at the origin of the cascading scenario of self-organization is very fast, its identifications was not yet possible. However, it was possible to obtain information on this mechanism starting from the experimentally proved fact that the CSCC, created by a cascading scenario of self-organization, strictly reveals the same features as those of a CSCC created by an intermittent scenario of self-organization [1,3,4]. Based on these experimental results, in the following we explain the creation of a CSCC by an electrical spark, considering the well-identified physical processes the successive development of which explains the generation of a CSCC by an intermittent scenario of self-organization.

## 2. Phenomenological model of a gaseous cell

Since life necessarily exists in the form of cells, the accumulation of charged particles in the form of a membranous boundary is the first structural requirement for the creation of a minimal prebiotic system. As mentioned, in a cold laboratory plasma in thermodynamic equilibrium, the premise for the formation of a gaseous membrane is the generation of a well localized non-equilibrium relatively high temperature plasma in a point where an electrical spark strikes the surface of a positively biased electrode [4]. Because of the differences in the mobility and thermal diffusivities of the electrons and the positive ions, the former are quickly collected by the positive electrode, so that a positive "nucleus" in the form of an ion-rich plasma appears. Acting as a gas anode, the potential of which depends on the positive electrode potential, the nucleus attracts, in the following phase of its evolution, electrons from the surrounding cold plasma. When the potential of the positive electrode, and implicitly of the gas anode, is so high that the accelerated electrons obtain kinetic energies sufficient to produce excitations and ionizations of the neutrals, the conditions for the self-assembly of a CSCC following an intermittent scenario of self-organization are realized [1,3-8]. It is well known that the excitation and ionization cross section functions depend on the kinetic energy of the electrons in such a way that the former suddenly increases for lower kinetic energies than the latter. Consequently, a net negative space charge populated with electrons that have lost their kinetic energy by excitation of neutrals at the specific energy levels is formed, in agreement with the equation $A + e_{fast} \rightarrow A^* + e_{slow}$. In this equation $A^*$ is an excited – but still neutral – atom, which, after about $10^{-8}$ s, returns into the ground state by emitting a photon with the energy $h\nu$. Therefore the region where the net negative charge is located appears as a luminous sheet that surrounds the positive nucleus [3]. Its extension depends on the scheme of the excitation levels of the respective gas atoms. The electrons, after having lost kinetic energy in excitation processes form a net negative space charge. This is dynamically maintained since the part of the accumulated electrons that disappear by recombination, diffusion etc. is continuously replaced by those electrons that have lost their momentum after excitations.

The appearance of a well-located net negative space charge represents the first phase in the pattern formation mechanism. Thus, the net negative space charge determines the location of the electric field in a relative small region at the border of the positive nucleus. This creates the premise for the following evolution sequence of the space charge into a CSCC. It is related to the electrons that have not produced excitations and, as a consequence of the acceleration in the electric field, they obtain sufficient kinetic energy to produce ionization in the nucleus according to the process $A + e_{fast} \rightarrow A^+ + 2e_{slow}$. Here $A^+$ is a single-ionized positive ion. Since the electrons that have produced ionizations and those resulting from these processes have low kinetic energies and are located in a relatively high electric field, formed between the net negative space charge and the positive electrode, the electrode quickly collects them. As a consequence, in this initial phase of the CSCC self-assembling process, there are two adjacent space charges of opposite sign in front of the anode: a positive one located in the nucleus and a negative one in the form of a sheet surrounding the nucleus. Between them the electrostatic forces act as long-range correlations, so that the adjacent space charge layers naturally associate in the form of a DL. So an electric field appears, within which the electrons obtain additional energies, so that the requirements necessary for the self-assembly mechanism of the DL are fulfilled at the boundary of the nucleus. The strength of this field depends on the densities of the two adjacent opposite net space charges, which in turn depend on the excitation and ionization rates and, implicitly, on the potential of the positive electrode. When the potential of the positive electrode is so high that the potential drop developed on the DL reaches the ionization potential of the gas, an instability will start to develop because a higher ion density in the nucleus causes an increase of the local electric field. As a consequence, the kinetic energy of the electrons accelerated towards the nucleus increases so that the ionization rate also grows. The result is an additional amount of positive ions that is added to the previous one. Thus, the local electric field and, implicitly, the ionization rate are further increased. In turn, the increase of the ionization rate produces an additional increase of the positive ion density and, consequently, a further grow of the electric field, and so on. As a result of this positive feedback mechanism the density of positive ions quickly grows in the region where the ionization cross section function suddenly increases (*i.e.* adjacent to the well-localized net negative space charge). In this way a *self-enhancement* mechanism for the production of positive ions working at the boundary of the nucleus governs the further evolution of the space charge configuration [8]. Once started at a given position, the sudden increase of the production rate of positive ions leads to an overall *"activation"* of this process. Therefore the self-enhancement of the production of positive ions alone is not sufficient to generate stable patterns in plasma. Their generation becomes, however, possible if the self-enhancement of the positive ion production is complemented by a mechanism able to act as a *"long-range inhibitor"* without impeding the incipient self-enhancement mechanism of the production of positive ions. The long-range inhibition mechanism is related to the creation of a negative space charge by neutral excitations acting as an "antagonist" to the positive one. Since also the excitation rates of neutrals depend on the kinetic energy of electrons, the increase of the electric field intensity, related to the self-enhancement of the production of positive ions, determines also the growth of the density of the adjacent negative space charge**.**

Simultaneously with the grow of the positive ion density, the local electric field increases, so that the region where the negative space charge forms by accumulation of those electrons that have lost their momentum by neutrals excitations is shifted away from the positive electrode. This "expansion" phase of the space charge configuration ceases because the production rate of positive ion cannot increase above a certain value when the neutral gas pressure is maintained constant. In the final phase of the CSCC evolution the negative space charge "balances" the positive space charge situated between it and the positive electrode. After this evolution the space charge configuration appears as a stable self-confined luminous, nearly spherical, gaseous body attached at the anode. Its self-assembling process governed by the electrostatic forces acting as correlations between the two adjacent net space charges of the DL, does not require additional energy since the transition occurs to a state characterized by a local minimum of the potential energy.

We note that in this stage of self-organization of the CSCC the DL from its border ensures its spatial stability by maintaining a local electric field that accelerates electrons at energies sufficient for producing within it the processes required for replacement of all of its components [1,8,9]. This becomes possible only when the transport of thermalized plasma electrons is ensured by the work done by the external dc power supply. A higher degree of self-organization of the CSCC appears when, after additional matter and energy injection, the CSCC transits into an open stationary state during which it undertakes a part of the work necessary for its self-existence. This is realized by a proper dynamics of the DL during which a rhythmic exchange of matter and energy with the surrounding environment takes place [6]. Both of these degrees of self-organization of the CSCC correspond to the "pre-natal" stages because their existence requires the presence of the external dc power supply.

The most interesting phenomenon observed in physical plasmas appears when the CSCC is created in low voltage thermionic arcs. In those plasma devices the CSCC emergence can also be explained considering an intermittent scenario of self-organization [11]. After its genesis the existence of the CSCC is possible in a free-floating steady state during which the DL at its border is subjected to a successive detachment and reformation processes. These phenomena reveal striking similarities with those observed when a CSCC is attached at the positive electrode of a plasma diode [6,12] or when it is created in an hf electric field [13]. Surprising is the fact that, during this free floating state, the mean potential of the nucleus of a CSCC exceeds the ionization potential of the gas also when the anode potential is much smaller than that. This proves that there exists a recharging mechanism of the nucleus of the CSCC by which its periodic discharging related to the transport of positive ions by the DLs, which periodically detach from its boundary, is compensated. This recharging mechanism, at present investigated in our laboratories, can be tentatively explained considering the experimental results that have proved that a moving DL is able to accelerate thermal electrons of the plasma, through which it propagates [14]. The recharging of the nucleus of the CSCC with positive ions becomes possible taking into account the Maxwellian energy distribution of the plasma electrons. Thus, obtaining kinetic energy by acceleration in the field of the moving (expanding) DL and also in the electric field of the net positive space charge (remaining in the nucleus after the DL detachment), the electrons reach the nucleus with energies sufficient to produce direct or stepwise ionizations. Since ionizations are accompanied by a heating process the temperature of the nucleus increases. This temperature increase is produced specially by the plasma electrons from the high energetic tail of the energy distribution. As a consequence a part of the electrons from the nucleus of the CSCC are ejected by thermal diffusion. After the ejection of electrons the potential of the nucleus attains again the value for which at its boundary a stable DL is self-assembled.

The periodicity of the process is related to the fact that after the detachment from the CSCC the spherical DL expands, so that the flux of electrons crossing it becomes too small to ensure its self-assembling process. Thus it decays. Thereafter the electrons, bounded at the DL space charge configuration, become free and are accelerated, as a bunch, towards the positive nucleus. Reaching the boundary of the nucleus, where a new DL is in a stable state, the flux of electrons increases the ionization rate so that the potential drop of the DL reaches the critical value for which its detachment process starts over. So, an internal working feedback mechanism ensures a rhythmic exchange of matter and energy between the CSCCC and the environment. Preliminary investigations performed in our laboratories seem to show that after the "birth" of the free-floating CSCC in a thermionic diode, its further existence does not require work from the external dc power supply. On the contrary, it is able to produce work by direct conversion of thermal energy in electric energy.

For high gas pressures, the described DL detachment and reformation processes take place in a relatively small region at the border of the free-floating CSCC. Conveniently, this region could be considered as playing the role of a "membrane" that protects the CSCC from the surrounding environment. Since the detachment of the DL involves the extraction and transport of positive ions from the nucleus of the CSCC to the surrounding plasma, a pressure difference appears between the former and the latter. Thus the periodic detachment of DLs from the boundary of the CSCC border implies a periodic "inhalation" of fresh neutrals into the nucleus. So, a CSCC also mimics the breathing process proper to all living systems.

By revealing the above described qualities, the free floating CSCC self-assembled in a thermionic diode is, to the best of our knowledge, the first autonomous complexity manufactured under controllable laboratory conditions able to ensure its survive by operations controlled by the DL at its boundary. Note that the excitation and ionization processes, the collective effects of which are at the basis of the operations performed by the DL are produced for relatively small values of electron kinetic energy.

For a biologist it could be of interest that the self-organization scenario, which explains the emergence of a CSCC, is essentially based on opposite space charge separation related to the symmetry breaking of specific quantum cross section functions. If a phenomenon, which is in principle similar to one acting in plasma, could also occur in chemical media, in which autocatalytic processes determine pattern formation, this could be a fascinating problem of further investigations. In this context we remember that, starting from the fact that symmetry breaking is a universally present phenomenon in biology, in his paper "Chemical Basis of Morphogenesis" Turing [15] proposed a mechanism for the generation of biological patterns. He considered a system of equations for chemical reactions, coupled by diffusion, which would deliver solutions that could break the symmetry of the initial state of a system. In the same context we remember that self-organization, related to symmetry breaking due to fluctuations in chemical systems with reaction and diffusion are regarded by Prigogine as a clue to the origin of life [16].

With respect to the cell models hitherto proposed, based on electronic circuits, constructed in order to simulate the electric activity of a biological system, the CSCC created by cascading self-organization, initiated by an electrical spark, reveals a electrical activity, potentially related to direct conversion of thermal energy in electric energy. A phenomenon that illustrates such a possibility is, in our opinion, the ball lightning, the occasional appearance of which proves the ability of Nature to create well located ordered space charge configurations [4]. As mentioned in the literature, the ball lightning deaggregation is accompanied by strong electrical oscillations. This seems to prove the presence of a steady state related to a dynamics of the DL at its boundary similar to that observed at a CSCC created in laboratory. The interpretation of the ball lightning as a "giant" cell seems to be justified if the described cascading scenario of self-organization actually stays at its origin [4]. Produced in an atmosphere that essentially differs from that presumably existent under pre-biotic Earth conditions, after its "birth" the evolution of a ball lightning ceases after a relatively short lifetime. This was, very probable, not the case when the CSCC self-assembly process was initiated under the prebiotic Earth conditions by a simple spark. Occurring in a medium, presumed to be a chemically reactive plasma, the possibility of a further evolution of a CSCC into the contemporary cell becomes a potentially possible alternative. In this context we remember that experiments performed in chemically reactive gases, simulating the prebiotic Earth conditions, have proved that micro-spheres (vesicles) protected from the environment by membranes able to support a potential drop, are self-assembled in electrical discharges [17].

As revealed by Nature, the creation of a living cell requires the self-assembly of a framework in the form of the cell membrane mainly constituted of lipids and proteins. The most important parts of this framework are the channels that, by a specific electric activity, control the matter and energy exchange between the nucleus of the cell and the surrounding medium. The force that maintains the ionic current through a channel has its origin in the electric potential produced by the gradients of the concentration of the different ion species inside and outside the cell. In the steady state of the cell, the local ionic influx compensates the ionic efflux. Why this gradient appears and acts, is today a challenging problem of Biology. In this context new information offered by biological observations have proved the presence of *pH* modulations. These are accompanied by the appearance of spatio-temporal patterns originating from a hypothetical self-organization scenario presumably related to a symmetry breaking mechanism [18]. For explaining the periodic pattern in the diffusion currents, induced by concentration gradients, a theory of electrodynamical instabilities was proposed linked to the specific properties of the membrane conductance [18].

An alternative explanation for the presence of spatio-temporal patterns in biological observations could be based on a self-organization mechanism as that one described in this paper. Such a mechanism becomes possible if a biological cell is the result of the evolution of a gaseous cell, formed by a cascading self-organization scenario in a chemical reactive medium, as that presumably present under primeval

Earth conditions. In that case the membrane of the cell must contain, in order to ensure its viability, channels able to maintain a local gradient of different ion species. This could be possible if at the ends of the channels "micro"-DLs are situated with qualities remembering their recent history. This means that the micro-DL preserves its initial ability to sustain and control an anomalous transport of matter and energy through the channel, by the described dynamics. It has obtained this ability during its creation under prebiotic Earth conditions. In this way the living state of a cell can be related to a mechanism able to explain the presence of periodic current patterns observed in the channels of its membrane. This mechanism can tentatively explain also the manner by which the pumping process is sustained in the channels of the cell membrane.

As mentioned, the CSCCs, created in plasma by self-organization, also reveal other interesting phenomena such as self-multiplication by division and exchange of information [19]. This latter behavior is realized by the emission of electromagnetic energy with an appropriate frequency by a CSCC during its steady (viable) state and its resonant absorption by another CSCC.

## 3. Conclusions

The identification of the physical causes of pattern formation in plasmas reveals the presence of a mechanism of self-organization that substantially advances the knowledge concerning the creation of complex systems in general. Based on phenomena such as local self-enhancement and long-range inhibition this scenario of self-organization shows striking similarities to those considered to lie at the heart of biological pattern formation [20]. Therefore for biologists the described self-organization scenario could be interesting accepting that this can explain the emergence of a primitive organism that could be a prerequisite condition for making possible the more complex chemical reactions. The space charges arrangement and the intrinsic nonlinear mechanism that ensure the "viability" of the CSCC are premises very probable necessary for a further chemical evolution into a organism revealing features as those proper to a contemporary cell. Additionally the described self-organization scenario suggest a new insight concerning the actual origin of various biological events as the morphogenesis, the polarization, the acquisition of nutriments and the mechanism by which electrical signals control the rhythm of a biochemical system. Clearly a major task of experimental studies are necessary to verify whether or not a spark produced at a positive electrode immersed in a chemical reactive plasma could initiate an "universal" self-organization process whose final product can eventually explain the origin of the life. Such experiments are at present performed in our laboratories.

**Acknowledgments.** The authors sincerely thank Professor Dr. Roman Schrittwieser from the University of Innsbruck, Austria for reading the paper and for many useful suggestions. This research was financially supported by the World Bank and by the Romanian Ministry of Education and Research under the Grant No.28199, CNCSIS code 47.